\documentstyle[prl,aps]{revtex}

\begin{document}

\draft
\twocolumn

\title{Information-theoretical meaning of quantum dynamical entropy}

\author{Robert Alicki}

\address{Institute of Theoretical Physics and Astrophysics, University
of Gda\'nsk, Wita Stwosza 57, PL 80-952 Gda\'nsk, Poland}

\date{\today}
\maketitle

\begin{abstract}
The theory of noncommutative dynamical entropy and quantum symbolic dynamics 
for quantum dynamical systems is analised from the point of view
of quantum information theory. Using a general quantum dynamical system as a communication
channel one can define different classical capacities depending on the character
of resources applied for encoding and decoding procedures and on the type of information
sources. It is shown that for Bernoulli sources the  entanglement-assisted
classical capacity, which is the largest one, is bounded from above by the quantum dynamical
entropy defined in terms of operational partitions of unity. Stronger results are proved for
the particular class of quantum dynamical systems -- quantum Bernoulli shifts. Different
classical
capacities are exactly computed and the entanglement-assisted one is equal to the dynamical
entropy in this case.  
\end{abstract}

\pacs{03.65.Fd, 89.70.+c }

\section{ Introduction}

The relations between the classical theory of dynamical systems and the theory of classical
communication channels are given by the Kolmogorov-Sinai construction of symbolic dynamics
and dynamical entropy (K-S entropy)[1]. One can expect that in the quantum domain the 
interrelations with information theory should
be much deeper. This is due to the fact that the quantum theory is a genuine statistical
and operational one and the most fundamental process - state preparation followed by  
measurement - possesses non-trivial information-theoretical meaning.  Indeed, take a tunable
device which prepars a quantum system in one of the states $\{\psi_1 , \psi_2,...\psi_m\}$ 
and use another apparatus to perform a measurement of the observable with possible values 
$\{a_1, a_2,...a_n\}$.
This can be seen as a single operation of an information channel with possible inputs
$\{1,2,...,m\}$ and outputs $\{1,2,...,n\}$. Quantum theory gives statistical 
predictions about the value of an output provided an input is given and optimalization 
of the transmitted information is a fundamental physical question.
\par
In the last decade a considerable progress in the theory of quantum communication channels
has been achieved [2,3]. However, most of the attention was concentrated on {\it memoryless
noisy channels}. They simulate physical systems essentially composed of noninteracting
subsystems (particles) with quantum noise acting independently on each of them. As a 
consequence, features of dynamics of the information carrier do not enter manifestly
the game. In order to investigate more complicated models of communication channels we use 
a different setting for sending classical information via quantum dynamical systems proposed
in [4]. In particular we expect relations between the speed of information transmission
through a channel and its chaotic properties characterized by quantum generalizations
of {\it K-S entropy}. In this scheme, not a presence of noise, but the way the 
perturbations of 
an initial state of the system propagate determines the efficiency of information 
processing. It is a well known fact used in modern control systems (e.g. aviation 
technology) that working in unstable (chaotic) regime is more efficient that in 
a stable one. A similar phenomenon should be visible in the quantum domain also.
\par   
In the theory of communication channels we are interested in asymptotic results valid
in the limit of infinitely long messages. Therefore the convenient mathematical description
involves infinite quantum systems similar to systems in thermodynamic limit considered
in statistical mechanics and quantum field theory. The corresponding mathematical formalism
is $C^*$-algebraic approach [5]. 
\par
\section{ Quantum dynamical systems in algebraic setting}

The approach to dynamical systems and dynamical entropy used here can be found in [6] 
together with a number of concrete examples and references to original papers and
alternative formalisms. 
\par
We assume that all bounded observables of our system generate a $C^*$-algebra 
${\cal A}$ with unit ${\bf 1}$. The discrete-time (reversible) dynamics is given in terms
of an automorphism
$\Theta$ acting on ${\cal A}$. By $\omega$ we denote a state on ${\cal A}$ invariant with
respect to $\Theta$. This state describes the reference (initial) state of the system,
for instance: ground state (e.g. vacuum state of the electromagnetic field), thermal 
equilibrium state, nonequilibrium stationary state (e.g. stream of particles), etc.
The triple $({\cal A},\Theta ,\omega)$ will be called a {\it quantum dynamical system}.
For infinite systems the $C^*-algebra$ ${\cal A}$ contains elements which do not correspond to observables
measured by any finite apparatus but rather describe properly defined limits of physical
observables. Therefore, it is necessary to consider a subalgebra ${\cal A}_0$ of physically
admisible observables called {\it local} or {\it smooth} subalgebra. The local algebra
should be invariant with respect to dynamics i.e. for $A\in{\cal A}_0,$ $\Theta(A)\in
{\cal A}_0$ too.  
\par
Ergodic properties of quantum dynamical systems are usually
defined in terms of system's reaction to external perturbations.
Any such perturbation can be realised by a {\it completely positive
unity preserving map} $\Lambda : {\cal A}\mapsto {\cal A}$. We restrict ourselves to
local and finite perturbations given by the formula
$$
\Lambda_{\bf X} (A) = \sum_{j=1}^k X_j^* A X_j
\eqno(1)
$$
where ${\bf X} = \{X_1,X_2...,X_k ; X_j\in {\cal A}_0 ;\sum_{j=1}^k X^*_j X_j = {\bf 1}\}$
is an {\it operational partition of unity}. 
\par
A completely positive map (1) perturbs the reference state $\omega$ yielding a new 
perturbed one which is defined in terms of the mean values 
$$
\omega^{\bf X} (A) = \sum_{j=1}^k \omega (X^*_j A X_j)\ ,\ A\in{\cal A}\ .
\eqno(2)
$$ 
One should notice that different partitions can produce the same completely positive
map. For two partitions ${\bf X,Y}$ we define a finer partition ${\bf X\circ Y}=
\{X_jY_l ; j=1,2,...,k , l=1,2,...,r\}$ and the corresponding completely positive
map $\Lambda_{\bf X\circ Y} = \Lambda_{\bf Y}\Lambda_{\bf X}$. Both, the set of partitions
${\cal P}({\cal A}_0)$ and the set of corresponding completely positive maps
${\cal M}({\cal A}_0)$ form semigroups with respect to
compositions and with a unity given by a trivial partition ${\bf I} = \{{\bf1}\}$. There are
important subsemigroups of partitions and maps :
\par
a)${\cal P}^b({\cal A}_0)$ and ${\cal M}^b({\cal A}_0)$ generated by bistochastic
partitions i.e. $\sum_{j=1}^k X_jX^*_j = {\bf 1}$ 
\par
b)${\cal P}^u({\cal A}_0)$ ${\cal M}^u({\cal A}_0)$ generated by unitary partitions
i.e. $X^*_jX_j = X_jX^*_j = \mu_j{\bf 1}; j=1,2,...,k$.
\par
Obviously, ${\cal P}^u({\cal A}_0)\subset{\cal P}^b({\cal A}_0)\subset{\cal P}({\cal A}_0)$ 
and ${\cal M}^u({\cal A}_0)\subset{\cal M}^b({\cal A}_0)\subset 
{\cal M}({\cal A}_0)$. The different subsemigroups correspond to the different
means used to perturb our system. Unitary partitions can be realized by external classical 
possibly random "potentials", bistochastic ones can involve interaction with  quantum
environment at infinite temperature (tracial) state, which displays some classical features, 
while general partitions need generic quantum ancillary resources. It is important that the
bistochastic maps does not decrease the entropy of the system. 

\subsection{Hilbert space representation}

It is often very convenient to use a canonical representation of a dynamical system 
$({\cal A},\Theta,\omega)$ in terms of:
\par
a) the Hilbert space ${\cal H}_{\omega}$ ,
\par
b) the representation of the algebra ${\cal A}$ in the algebra of bounded operators
$B({\cal H}_{\omega})$ i.e. any element $A$ of ${\cal A}$ is
represented by an operator ${\hat A }$ and the map $A\mapsto {\hat A}$ preserves
the algebraic structure,
\par
c) the state $\omega$ is represented by the normalized Hilbert
space vector $|\Omega>\in {\cal H}_{\omega}$ such that 
$\omega (A) = <\Omega, {\hat A} \Omega>$,
\par
d) the dynamical automorphism $\Theta $ is represented by the unitary operator ${\cal U}$ ,
${\cal U}|\Omega> = |\Omega>$ and for $B= \Theta (A) ,\  {\hat B} 
= {\cal U}^* {\hat A}{\cal U}\equiv {\hat \Theta}({\hat A})$. It is useful to define 
the Schr\"odinger picture
of the dynamics in the Hilbert space representation by a transposed map
$$
{\hat {\Theta}}^T({\hat {\rho}}) =  {\cal U} {\hat{\rho}}{\cal U}^*\ .
\eqno(3)
$$ 

The Hilbert space ${\cal H}_{\omega}$ can be identified with 
the algebra ${\cal A}$ equipped with  the scalar product
$<A,B>\ =\ \omega (A^*B)$
modulo the equivalence
relation: $A\equiv B$ if and only if $\omega
[(A-B)^*(A-B)]=0$.
The element defined by the unity ${\bf 1}$ in ${\cal A}$ is exactly our normalized vector 
$\Omega$.
Any element $A$ of the algebra ${\cal A}$ is
represented by the operator ${\hat A }$ which is defined by the
left multiplication. Operators corresponding to right multiplication form an algebra of
observables of a "minimal environment" (ancilla). 
\par
The formalism of above, called in the mathematical literature
GNS representation, has been rediscovered by physicists 
under the names of the Liouville space approach [7], thermofield formalism [8]
or in the context of quantum information theory as "state purification by ancilla" [2].
Its physical meaning
for finite systems is clear: for a system being in a mixed state we reconstruct its 
minimal dilation described by a pure entangled state which reproduces the original state 
as a reduced marginal one.     
\par
Completely positive maps discussed above act in GNS representation on the whole operator
algebra $B({\cal H}_{\omega})$
$$
{\hat{\Lambda}}_{\bf X} (B) = \sum_{j=1}^k {{\hat X}_j}^* B {\hat X}_j\ ,\ 
B\in B({\cal H}_{\omega})
\eqno(4)
$$
and the perturbed state $\omega^{\bf X}$ is represented by the density matrix
$$
{\hat\rho} [{\bf X}] = {{\hat {\Lambda}}_{\bf X}^T (|\Omega><\Omega|) =
\sum_{j=1}^k |{\hat X}_j}\Omega><{\hat X}_j\Omega|
\eqno(5)
$$ 
where ${{\hat {\Lambda}}_{\bf X}}^T$ is the (GNS) Schr\"odinger picture version of the 
Heisenberg picture map ${\hat\Lambda}_{\bf X}$. 
\subsection {Quantum dynamical entropy}

We briefly sketch the theory of quantum dynamical entropy defined in terms of
operational partitions of unity. The basic object in this approach is the following 
multi-time correlation matrix generated by the partition ${\bf X} = \{X_1, X_2,...,X_k\}$ 
$$
\rho[{\bf X}^n]_{i_1,...,i_n;j_1,...,j_n}  = 
$$ 
$$
\omega\bigl(X^*_{j_1}\Theta(X^*_{j_2})...\Theta^{n-1}(X^*_{j_n})
\Theta^{n-1}(X_{i_n})...\Theta (X_{i_2})X_{i_1}\bigr)\ .
\eqno(6)
$$
$\rho[{\bf X}^n]$ is a positively
defined, $k^n \times k^n$ complex-valued matrix with a trace
equal one. Therefore, the sequence $\{\rho[{\bf X}^n]; n=1,2,3,...\}$ can be treated
as a consistent family of reduced density matrices which describes
the state of a one-sided chain of quantum "spins". To any spin at a given 
site corresponds
a $k$-dimensional Hilbert space and $\rho[{\bf X}^n]$ is a mixed state
of $n$ spins located at the sites $\{0,1,...,n-1\}$. Then a single step
of the evolution translates into the right shift on the spin chain and
we obtain a {\it quantum symbolic dynamics }.
\par
The von Neumann entropy of the density matrix $\rho[{\bf X}^n]$ measures
an amount of information encoded in the multitime correlations:
$$ 
S(\rho[{\bf X}^n]) = -{\rm tr} \bigl(\rho[{\bf X}^n]\ln \rho[{\bf X}^n]\bigr)
\eqno(7)
$$
The entropy of the partition $h[\omega,\Theta,{\bf X}]$ is defined as a limit
$$ 
h[\omega,\Theta,{\bf X}]  = \limsup_{n\to\infty}{1\over n} S(\rho[{\bf X}^n])
\ .
\eqno(8)
$$
Finally, the dynamical entropy of $\Theta$ is a supremum over all physically
admisible (local) partitions
$$ 
h[\omega,\Theta,{\cal A}_0] = \sup_{{\bf X}\in{\cal P}({\cal A}_0)}
 h[\omega,\Theta,{\bf X}]\ .
\eqno(9) 
$$ 
Restricting the supremum to subsemigroups ${\cal P}^u({\cal A}_0)$ or
${\cal P}^b({\cal A}_0)$ we obtain corresponding restricted dynamical entropies satisfying
the obvious inequality
$$
h^u[\omega,\Theta,{\cal A}_0]\leq h^b[\omega,\Theta,{\cal A}_0] 
\leq h[\omega,\Theta,{\cal A}_0]\ . 
\eqno(10)
$$
The equivalent expression for (8) can be obtained in the GNS representation
$$ 
S(\rho[{\bf X}^n]) = S({\hat \rho}[{\bf X}^n])
\eqno(11)
$$
where (see eqs (5)(6)) 
$$
{\hat \rho}[{\bf X}^n] = 
[{\hat \Theta}^T{\hat \Lambda}^T_{\bf X}]^n(|\Omega><\Omega|)\ .
\eqno(12)
$$ 
The formula of above suggests a new 
interpretation of $S(\rho[{\bf X}^n])$ as the entropy of the density matrix
obtained by repeated measurements performed on the evolving system $+$ ancilla.
\par
It has been proved that for classical systems the scheme of above reproduces the 
standard Kolmogorov-Sinai entropy, and a for a number of infinite quantum systems
the dynamical entropy (9) has been computed. Moreover, in the known examples all three 
entropies (10) coincide. Although, strictly speaking, the dynamical entropy exists
only for classical or infinite quantum systems the n-dependence of the entropy 
$S(\rho[{\bf X}^n])$ provides interesting informations about "quantum chaos" in finite 
quantum systems as well.
\par
There exist other, nonequivalent definitions of quantum dynamical entropy among them
CNT-entropy is the most developed one [9]. Preliminary results on its 
information-theoretical meaning can be found in [10].

\subsection {Quantum Bernoulli shifts}
The simplest example of infinite quantum dynamical system is a quantum Bernoulli shift.
Consider an infinite collection of the
identical quantum systems ("spins") attached to the sites
of 1-dimensional lattice labeled by the integers ${\bf Z}$.
The single site algebra is a $d\times d $ matrix algebra ${\bf M}_d$ and 
${\cal A}_{[-n,n]}$ denotes the algebra localized on $[-n,n]$ and given by
a suitable tensor product of ${\bf M}_d$. 
The local algebra of observables ${\cal A}_0 = 
\bigcup_{n\in {\bf N}}{\cal A}_{[-n,n]}$ can be completed to
a $C^*$- algebra ${\cal A}$ of {\it quasi-local observables}. 
The discrete time dynamics $\Theta$ is given
by a shift to the right which is an automorphism on ${\cal A}$
leaving ${\cal A}_0$ invariant.
The state $\omega$ of the considered system is a product state
$\otimes_{\bf Z}\rho$ where $\rho$ is a single-site state given by
a $d\times d$ density matrix. Obviously, $\omega$ is shift invariant. 
\par
One can easily compute the dynamical entropies (9)(10) for the quantum
Bernoulli shift which are equal
$$
h^u[\omega,\Theta,{\cal A}_0]= h^b[\omega,\Theta,{\cal A}_0] 
= h[\omega,\Theta,{\cal A}_0]= S(\rho) + \ln d \ . 
\eqno(13)
$$
To prove it one can notice that the RHS of eq.(13) is an upper bound for any
$h[\omega,\Theta,{\bf X}]$ , ${\bf X}\in {\cal P}({\cal A}_0)$ due
to a general inequality
$$
S(\sigma [{\bf X}])\leq S(\sigma) + \ln N
\eqno(14)
$$
where $\sigma$ is a density matrix on an $N$- dimensional Hilbert space,
${\bf X}$ is an arbitrary partition of unity and $ \sigma [{\bf X}]_{ij} =
{\rm tr}(\sigma X^*_jX_i)$. This bound is reached for a local, single-site 
unitary partition
$$
{\bf W} = \{d^{-1}W(k,l); k,l = 1,2,...,d\}
\eqno(15)
$$
where $W(k,l)$ are {\it discrete Weyl operators} defined in terms of the basis
$\{|e_m> ; m=1,2,...,d\}$ of eigenvectors of $\rho$ by the formula
$$
W(k,l)|e_m> = \exp (i2\pi k/d) |e_{m\oplus l}>\ , 
$$
$$
m\oplus l= m+l \ ({\rm mod}\ d)\ .
\eqno(16)
$$ 
\section{Communication channel with classical input and output}

We consider a model of communication channel
for which the input and output are strings of letters and
the physical carrier of information is a quantum dynamical
system described in the $C^*$-algebraic language by $({\cal A},{\cal A}_0,
\Theta,\omega)$ as in the previous Section.
\subsection{Input and output}
A given input
message of the length $n$ is a sequence $\alpha_1,\alpha_2,... \alpha_n $ of letters 
which belong to a certain alphabet identified with $\{1,2,...,r\}$. Any letter $\alpha$
is transmitted by means of a perturbation of the reference
state $\omega $ by a completely positive map $\Lambda_{\alpha}\in {\cal M}({\cal A}_0)$.
This encoding procedure will be shortly denoted by ${\bf {\Lambda}}$.
We can restrict possible perturbations to entropy increasing ones i.e.
$\Lambda_{\alpha}\in {\cal M}^u({\cal A}_0)$ or
$\Lambda_{\alpha}\in {\cal M}^b({\cal A}_0)$.
Two consecutive perturbations of the state $\omega$ are always separated by 
the action of the dynamics $\Theta$. This can be regarded as a definition of a letter which
is the basic unit of the message sent during the single evolution step (unit of time). 
Therefore for a n-letter message we have the 
corresponding completely positive perturbation
$$
{\bar\alpha}\equiv(\alpha_1,\alpha_2,... \alpha_n )\mapsto  \Lambda_{\alpha_1}\Theta 
\Lambda_{\alpha_2}\Theta\cdots\Lambda_{\alpha_n}\Theta\ .
\eqno(17)
$$
It is convenient to use the Hilbert space (GNS) representation to associate with 
a given massage ${\bar \alpha} = (\alpha_1,\alpha_2,... \alpha_n )$ a density matrix
${\hat\rho}({\bar \alpha})$ acting on the Hilbert space ${\cal H}_{\omega}$
which can be written using the notation (6)(12)(17) as
$$
 {\hat\rho}({\bar \alpha})=  {\hat\Theta}^T{\hat\Lambda}_{\alpha_n}^T\cdots 
{\hat\Theta}^T{\hat\Lambda}_{\alpha_2}^T{\hat\Theta}^T{\hat\Lambda}_{\alpha_1}(|\Omega><\Omega|)\ .
\eqno(18)
$$

Receiving of a message is realized by performing a measurement of the suitable decoding
observable ${\bf D}$ with possible outcomes $(\delta_1,\delta_2,...\delta_m)$.
Here ${\bf D} = \{D_1,D_2,...D_m ; D_k\in B({\cal H}_{\omega}), D_k \geq 0, \sum_{k=1}^m
D_k = {\bf 1}\}$ is a {\it generalized observable } (or "fuzzy observable"). Choosing $D_k$
from the whole $B({\cal H}_{\omega})$ means that we are able to exctract the information
encoded in entanglement of the system with its environment. If we can perform the 
measurements on  the dynamical system only we put $D_k\in {\cal A}$ identifying any element
of ${\cal A}$ with its operator representation in $B({\cal H}_{\omega})$.
\par
The basic quantity is  the conditional output probability
$ P({\bar\alpha} |\delta_j )$ which gives the probability of recording the output
$\delta_j$ under the condition of the input message ${\bar\alpha}$ [9]
$$ 
P({\bar\alpha} |\delta_j) = {\rm tr}({\hat\rho}({\bar \alpha})D_j)\ .
\eqno(19)
$$
Having a given input probability distribution $p_{in} = \{p_{in}({\bar\alpha})\}$ 
we can define the output probability
distribution
$$p_{out}(\delta_j) = \sum_{\bar\alpha} p_{in}({\bar\alpha}) P({\bar\alpha} |\delta_j) 
\eqno(20)
$$
and the input-output probability distribution
$$ p_{in,out}({\bar\alpha} ,\delta_j) = p_{in}({\bar\alpha}) P({\bar\alpha} |\delta_j)\ .
\eqno(21)
$$
The standard definition of the {\it amount of 
transmitted information} is given in terms of Shannon entropies $S(p) = -\sum p_k\ln p_k$
[11]
$$ 
I(p_{in},{\bf\Lambda},{\bf D}) = S(p_{in}) + S(p_{out}) - S(p_{in,out}) 
$$
$$
= S(p_{out}) - \sum_{\bar\alpha} p_{in}(\alpha) S(P({\bar\alpha} |\cdot))
\eqno(22)
$$
and satisfies the following inequalities
$$ 0\leq I(p_{in},{\bf\Lambda},{\bf D})\leq \min \{S(p_{in}),S(p_{out})\}\ .
\eqno(23)
$$
The Holevo-Levitin inequality [11] provides an upper 
bound on $I(p_{in},{\bf\Lambda},{\bf D})$ which is
independent of the choice of an output device
$$
I(p_{in},{\bf\Lambda},{\bf D})  \leq S\Bigl( \sum_{\bar\alpha} p_{in}({\bar\alpha}) 
{\hat\rho}({\bar\alpha})\Bigr)
- \sum_{\bar\alpha} p_{in}({\bar\alpha}) S({\hat\rho}({\bar\alpha}))\ .
\eqno(24)
$$
\subsection {Channel capacities}
The important quantity which characterizes the efficiency of a
communication channel is its capacity. In our case it will be an averaged amount 
of classical information, transmitted per unit of time, maximized over definite
sets of information sources, encoding and decoding procedures and calculated in the
limit of infinitely long input messages
$$
{\cal C} = \sup_{\{p_{in}\},\{\bf\Lambda\},\{\bf D\}}\Bigl\{\limsup_{n\to\infty}{1\over n}\,
I(p_{in},{\bf\Lambda},{\bf D})\Bigr\}\ .
\eqno(25)
$$
In the following we shall discuss several cases of capacity:

a)The entanglement-assisted classical capacity $C_E$ [12] which corresponds to the supremum 
taken over all information sources, arbitrary encoding procedures ${\bf\Lambda}\subset 
{\cal M}({\cal A}_0)$  
and arbitrary decoding observables ${\bf D}\subset B({\cal H}_{\omega})$ .

b)The ordinary classical capacity $C$ and its rectricted versions $C_u$, $C_b$ corresponding
to the supremum over all information sources, arbitrary decoding observables of the system 
alone, i.e. ${\bf D}\subset {\cal A}$, and encoding procedures involving completely positive
perturbations from ${\cal M}({\cal A}_0)$, ${\cal M}^u({\cal A}_0)$ and ${\cal M}^b({\cal A}_0)$
respectively.

c)The capacities of above restricted to Bernoulli sources, i.e. the information 
sources  with  product probability measures 
$$ p_{in}(\alpha_1,\alpha_2,...,\alpha_n) =
p(\alpha_1) p(\alpha_2)...p(\alpha_n)
\eqno(26)
$$
and denoted by $C^0_E,C^0,C^0_u,C^0_b$.
\par
The definitions of above imply obvious inequalities
$$
C^0_u \leq C^0_b\leq C^0\leq C^0_E\   ,\ C_u \leq C_b\leq C\leq C_E ,
$$
$$
C^0_E\leq C_E\ ,\ C^0_u \leq C_u\  ,\ C^0_b \leq C_b\ ,\  C^0 \leq C\ .
\eqno(27)
$$ 
The dynamics $\Theta$ of the system is reversible and therefore noise is not explicitly
present in this scheme. There are several possibilities to introduce noise in our setting.
The first, natural one, seems to be replacing an authomorphism $\Theta$ by a completely 
positive dynamical map. However, this would produce capacities typically 
equal to zero because 
the errors accumulate with a number of time steps except the situation where a proper 
scaling of noise with $n$ is introduced. Another 
possibility consists in putting extra conditions on decoding observables ${\bf D}$, 
assuming that ${\bf D}\subset {\cal B}$ where ${\cal B}$ is a proper subalgebra of ${\cal A}$
or $B({\cal H}_{\omega})$. A certain type of background noise appears
in the case of capacities $C_u , C_b$ or $C^0_u , C^0_b$ due to a mixed reference  
state $\omega$ which cannot be locally purified by entropy increasing perturbations 
(see Section IV).
\subsection { Dynamical entropy bound}

We prove our first result which provides the relation between ergodic properties
of the channel treated as a dynamical systems and its entanglement-assisted capacity
for the case of Bernoulli sources.
\par
{\it Theorem\ 1 } For any quantum dynamical system 
$$
C^0_E\leq h[\omega,\Theta,{\cal A}_0]\ .
\eqno(28) 
$$  
The proof follows from the Holevo-Levitin inequality (24) and the definitions
(8)(9). For a Bernoulli source and a given encoding ${\bf\Lambda}$ there exists a
partition of unity ${\bf X}$ such that $\Lambda_{\bf X} = 
\sum_{\alpha}p(\alpha) \Lambda_{\alpha}$
Therefore, for any message of length $n$ (see eqs(12)(18))
$$
\sum_{\bar\alpha} p_{in}({\bar\alpha}) {\hat\rho}({\bar\alpha})=
{\hat \rho}[{\bf X}^n] = [{\hat \Theta}^T{\hat \Lambda}^T_{\bf X}]^n(|\Omega><\Omega|)
\eqno(29)
$$
and applying (9)(24)
$$
\limsup_{n\to\infty} {1\over n}I(p_{in},{\bf\Lambda},{\bf D})\leq
h[\omega,\Theta,{\bf X}] \leq h[\omega,\Theta,{\cal A}_0]\ .  
\eqno(30)
$$  
The natural questions arise, how tight is this bound and whether it is possible
to prove it for more general sources. This will be discussed in the next Section
for quantum Bernoulli shifts.

\section{Capacities for Quantum Bernoulli Shifts}

Perturbations of the reference state for quantum Bernoulli shifts propagate in a very 
simple way what allows to prove much stronger results than those given by (27)(28).
\par
{\it Theorem\ 2} For a quantum Bernoulli shift the following equalities hold
$$
C^0_u = C^0_b =C_u =C_b = \ln d - S(\rho)\ ,
\eqno(31)
$$
$$
C^0 =C = \ln d\ ,
\eqno(32)
$$
$$
C^0_E =C_E =  \ln d + S(\rho)\ .
\eqno(33)
$$
The interpretation of these results is quite obvious. The nonzero single-site entropy
$S(\rho)$ can be an obstacle (noise) or an asset depending on the control we have
of the system and its environment. Assume first, that we have no acces to environment.
Then, if we can use  entropy increasing perturbations only, $S(\rho)$ is an amount
of noise which reduces the capacity of the channel. Applying arbitrary encoding
with the help of ancillary resources we can reach a capacity $\ln d$. 
\par
On the other hand, if we can 
control the environment, represented here as an ancillary spin chain with
the prior entanglement for any pair spin-ancilla, $S(\rho)$ becomes an amount of entaglement
per site which improves the capacity. This is exactly the idea of 
{\it quantum dense coding} [2].
Moreover, $C^0_E$ reaches its upper bound (28) and 
the Bernoulli sources are optimal for all studied examples of capacities. One can expect
that the bound (28) is tight and the Bernoulli sources are optimal for a larger class
of quantum dynamical systems at reference states satisfying certain clustering properties 
with respect to dynamics. Finally, one should notice that for the quantum Bernoulli shift
the CNT-entropy is equal to $S(\rho)$. 

\subsection {Proof of Theorem 2}

In the first part of the proof we use again Holevo-Levitin inequality (24)and the fact that 
all perturbations of the state are strictly local and their propagation is given simply 
by a shift. Hence, for a given encoding ${\bf\Lambda}$ all completely positive
maps are localised on the sites in a certain interval $[-l , l]$. After $n$ time steps
the total perturbation is localised in $[-l, l+n]$ and the perturbed state on the quasilocal
algebra $\bigotimes_{\bf Z} M_d$ can be replaced by a local density matrix
$$
\rho ({\bar\alpha})= \Lambda^T ({\bar\alpha})(\otimes_{[-l,l+n]}\rho)
\eqno(34)
$$
where $\Lambda^T ({\bar\alpha})$ is a total perturbation map in Schr\"odinger picture
(not to be confused with GNS representation (18)!). Applying now inequality (24)
for the spin system living on the interval $[-l,l+n]$
$$
I(p_{in},{\bf\Lambda},{\bf D}) \leq S\Bigl(\sum_{\bar\alpha}p_{in}({\bar\alpha}) 
\rho ({\bar\alpha})\Bigr) -  \sum_{\bar\alpha}p_{in}({\bar\alpha}) 
S(\rho ({\bar\alpha}))   
\eqno(35)
$$
we obtain for arbitrary perturbations
$$
I(p_{in},{\bf\Lambda},{\bf D}) \leq (n+2l+1)\ln d
\eqno(36)
$$
while for the entropy increasing ones
$$
I(p_{in},{\bf\Lambda},{\bf D}) \leq (n+2l+1)(\ln d - S(\rho))\ . 
\eqno(37)
$$
To obtain a bound useful for $C_E$ we use a GNS representation and the bound
$$
I(p_{in},{\bf\Lambda},{\bf D}) \leq S\Bigl(\sum_{\bar\alpha}p_{in}({\bar\alpha}) 
{\hat\rho}({\bar\alpha})\Bigr)\ .
\eqno(38)
$$
For any $n$ there exists a partition of unity ${\bf Y}_n$ which is generally not a 
composition of $n$ partitions like ${\bf X}^n$ in (6) but nevertheless is localised
on the interval $[-l, l+n]$ such that
$$
\sum_{\bar\alpha}p_{in}({\bar\alpha}) {\hat\rho}({\bar\alpha})
={\hat\rho}[{\bf Y}_n]\ .
\eqno(39)
$$
Then using (24) and the general bound (14)
$$
I(p_{in},{\bf\Lambda},{\bf D}) \leq S\Bigl(\sum_{\bar\alpha}p_{in}({\bar\alpha}) 
{\hat\rho}({\bar\alpha})\Bigr)\ 
$$
$$
= S({\hat\rho}[{\bf Y}_n]) = S({\rho}[{\bf Y}_n]) \leq (n+2l+1)(S(\rho)+ \ln d)\ .
\eqno(40)
$$ 
The proof of the upper bounds is completed by dividing both sides of (36)(37)(40) by $n$ ,
taking limit $n\to\infty$ and proper suprema over $p_{in}, {\bf\Lambda}$ and ${\bf D}$.
\par
In the second part of the proof we show that the upper bounds are reached choosing
proper Bernoulli sources, single-site encoding perturbations and suitable decoding
observables. In this case $p_{in}({\bar\alpha}) = p(\alpha_1)p(\alpha_2)\cdots p(\alpha_n)$
and $\rho({\bar\alpha}) = \rho(\alpha_1)\rho(\alpha_2)\cdots \rho(\alpha_n)$ what is exactly
the setting of the Holevo-Schumacher-Westmoreland theorem [13] which may be formulated as
follows.
\par
{\it Theorem\ 3} Take a Bernoulli source and a single-site encoding as above. Then,
by a suitable choice of  a decoding observable the 
asymptotic amount of transmitted information per unit of time can be arbitrarily close to
the Holevo-Levitin bound 
$$
S\bigl(\sum_{\alpha} p(\alpha)\rho(\alpha)\bigr) -\sum_{\alpha} p(\alpha)
S\bigl(\rho(\alpha)\bigr)\ .
\eqno(41)
$$
\par  
It remains to compute the bound (41) for different schemes corresponding to the capacities
$C^0, C^0_u $ and $C^0_E$ respectively.
\par
For $C^0$ we take $d$ letters with a priori probabilities $1/d$ and the single-site 
perturbations
$$
\Lambda^T_{\alpha}(\sigma) = {\rm tr}(\sigma)|e_{\alpha}><e_{\alpha}|\ ,\
\sigma\in {\bf M}_d
\eqno(42)
$$
where $\{|e_{\alpha>}\}$ is a basis for a single spin. The bound (41) is obviously equal to
$\ln d$.  
\par
For $C^0_u$ we take equally distributed $d^2$ letters with unitary single-site encoding
given by the discrete Weyl operators $W(l,k)$ (15). Then using the fact that for any single spin
matrix $\sigma$
$$
{1\over d^2}\sum_{k,l=1}^d W(k,l)\sigma  W(k,l)^* = {\rm tr} (\sigma){1\over d} {\bf 1}
\eqno(43)
$$
we obtain the bound (41) equal to $\ln d - S(\rho)$.
\par
To reach the bound for $C^0_E$ we consider a purification of Bernoulli shift with 
a pure single-site reference state of spin - ancilla
$$
{\tilde\rho} = \sum_{j=1}^d {\sqrt\lambda_j}|e_j>\otimes |e'_j>
\eqno(44)
$$
being a purification of $\rho = \sum_{j=1}^d {\lambda_j}|e_j><e_j|$ .
Taking again equally distributed $d^2$ letters with unitary single-site encoding
given by the unitary operators $W(l,k)\otimes{\bf 1}$ we reach the bound $\ln d + S(\rho)$.

\acknowledgments
The work is supported by the Grant 2P03B042 of the Polish Committe for Scientific
Research.

\end{document}